\documentclass[%
 reprint,
 amsmath,amssymb,
 aps,
floatfix,
]{revtex4-2}

\usepackage{graphicx}
\usepackage{dcolumn}
\usepackage{bm}

\begin{document}
\preprint{APS/123-QED}

\title{Inhibited spontaneous emission of quantum dots weakly coupled to off-resonant silver nanoplatelets and silver nanowires}
\author{Harshavardhan R. Kalluru\textsuperscript{1}}
\email{kallurureddy@iisc.ac.in}
\affiliation{Department of Physics, Indian Institute of Science Bengaluru, 560012, India.\textsuperscript{\textup{1,2,3,4}}}
\author{Binita Tongbram\textsuperscript{2}}
\affiliation{Department of Physics, Indian Institute of Science Bengaluru, 560012, India.\textsuperscript{\textup{1,2,3,4}}}
\author{Ashish Biswas\textsuperscript{3}}
\affiliation{Department of Physics, Indian Institute of Science Bengaluru, 560012, India.\textsuperscript{\textup{1,2,3,4}}}
\author{Jaydeep K. Basu\textsuperscript{4}}
\affiliation{Department of Physics, Indian Institute of Science Bengaluru, 560012, India.\textsuperscript{\textup{1,2,3,4}}}
\date{\today}
\begin{abstract}
Spontaneous emission (SE) rate of any light emitters directly scales with the locally available modes for photons. The emission rate can be modified, by changing the dielectric environment of light emitters. Generally cavities with modes in resonance to light emission frequency, are used to amplify the light emission rate. The Fermi golden rule predicts that if the cavity modes are off-resonant to the emission frequency, then the SE rate is suppressed. In this study, we demonstrate that the SE of colloidal alloyed quantum dots is inhibited by coupling them to chemically synthesized Silver nanowires and Silver nanoplatelet systems. The silver nanoplatelet and silver nanowire plasmonic resonance modes are in ultraviolet and infrared regions of the electromagnetic spectrum. The quantum dots emit in visible region of light. This off-resonant weak coupling of emitters and cavities results in emission rate suppression and is quantified by time resolved photoluminescence (TRPL) measurements. TRPL decay profiles show that the emission rate can be suppressed by coupling self assembled quantum dot monolayers to a single silver nanoplatelet and a single silver nanowire respectively.
\end{abstract}
\maketitle
\section{Introduction}
The spontaneous emission (SE) rate changes with the available cavity modes in the vicinity of emitters. This was first reported by E.M. Purcell.\cite{10.1103/PhysRev.69.674.2} The SE rate is directly proportional to the local photonic density of states (PDOS). So in presence of cavity modes which are resonant to emission frequency, the SE rate is enhanced relative to vacuum emission rate.\cite{10.1021/nl0715847} If the cavity modes are off resonant, then the SE rate is inhibited relative to vacuum emission rate.\cite{10.1021/nn5017555} The PDOS can be quantified and measured as the ratio of the spontaneous radiaitive emission rate near a cavity ($\Gamma_\textup{c}$) relative to spontaneous radiative emission rate in vacuum ($\Gamma_\textup{o}$). The ratio is known as Purcell factor. Theoretically, the Purcell factor (F), for the cavity modes resonant to the quantum emission frequency\cite{PhysRevLett.110.237401} is given by
\[\textup{F}=\frac{3\lambda_o^3}{4\pi}\cdot\Big(\frac{Q}{V}\Big)\tag{1}\]
where $Q$ is the cavity quality factor, $V$ is cavity mode volume and $\lambda_{o}$ is the cavity mode resonance wavelength. If the cavity modes are off-resonant to the quantum dot emission frequency, the Purcell factor is given by
\[\textup{F}=\frac{3\lambda_o^3}{16\pi}\cdot\Big(\frac{1}{QV}\Big)\tag{2}\]

The spontaneous emission rate ($\Gamma$) has two components, one is radiative decay rate ($\Gamma_{\textup{r}}$) and the other one is non-radiative decay rate ($\Gamma_\textup{nr}$). As the radiative emission rate is the reciprocal of the spontaneous life time ($\tau$) and the radiative part of the Purcell factor can be experimentally estimated by time resolved photoluminescence (TRPL) measurements. 
\[\Gamma=\Gamma_\textup{r} + \Gamma_\textup{nr}\tag{3}\]
For silver plasmonic cavities, the absorption of light through non-radiative decay occurs via four mechanisms.\cite{10.1038/nnano.2014.310} They are classified into phonon mediated absorption, absorption mediated by electron-electron scattering, direct surface plasmon scattering mediated absorption and inter-band absorption processes. The direct experimental measurement of non-radiative part of Purcell effect, involves accounting absorption for each of these four processes and is beyond the scope this manuscript. So, only the radiative part of Purcell effect is discussed.

\[\Gamma_\textup{r}=\frac{\Gamma_\textup{c}}{\Gamma_\textup{0}}=\frac{\tau_\textup{o}}{\tau_\textup{c}}\tag{4}\]

 Here $\Gamma_{o}$ and $\Gamma_{c}$ are the radiative decay rate of emitters in vacuum and near the cavity. $\tau_\textup{o}$ and $\tau_\textup{c}$ are the corresponding lifetimes of the emitters in vacuum and near the cavity. If the cavity mode is resonant to the emission, then the $\textup{F}> 1$, which means the radiative emission rate is enhanced . If the cavity mode is off-resonant, then the $\textup{F}< 1$, which means the radiative emission rate is inhibited. The resonant and off-resonant emission rate modification is a consequence of weak coupling between the emitter and cavity modes.

In this study, CdSe-ZnS alloyed quantum dots (AQDs) are used as emitters. The plasmonic cavities used are Silver nanowires (AgNWs) and Silver nanoplatelets (AgNPLs). The self-assembled AQD layers are transferred onto the cavities and the coupled systems are studied. The position of plasmonic dipolar modes (DM) of the AgNPLs and AgNWs, depends upon the aspect ratio (a) of these systems. The aspect ratio\cite{10.1002/smll.200801480} is defined as the ratio of lateral size to thickness of a nanostructure. In the case of AgNPLs, the ratio of triangular edge length to AgNPL thickness is considered as aspect ratio. In the case of AgNWs, the ratio of wire length to wire thickness is considered as the aspect ratio. Typically for the synthesized AgNWs in this study, aspect ratio is 241. AgNPLs are synthesized in three aspect ratios of 3, 261 and 551. 

AgNWs have localized surface plasmon resonances along longitudinal and transverse directions of the wire.\cite{PhysRevLett.95.257403} Similarly triangular AgNPLs have in-plane oriented plasmonic dipolar modes (DM) and quadrupolar modes (QM).\cite{10.1021/nl061286u,10.1021/nl0340475,10.1002/adfm.200800233} It is well established that for silver nanowires, the longitudinal surface plasmon resonance (LSPR), moves from visible to infra red region with increasing aspect ratio.\cite{10.1021/jp0107964,10.1021/jz4005015} The absorbance peaks of LSPR peaks are in infra red region and can be measured by electron energy loss spectroscopy (EELS).\cite{PhysRevLett.110.066801} The AgNPL dipolar plasmonic resonance also shifts to higher wavelengths with increasing aspect ratio.\cite{10.1039/C6CP00953K,10.1002/adfm.200500667} So considering the large aspect ratios of silver nanoplatelets and silver nanowires in this study, all the plasmonic mode resonances are off resonant relative to the AQD emission.

\section{Experimental Methods}
The AQDs are synthesized by protocol mentioned in\cite{10.48550/arxiv.2212.13510}. The AgNPLs and AgNWs are synthesized and cleaned by protocols mentioned in appendices A and B respectively. The synthesized nanoplatelets and nanowires are imaged and characterized by scanning transmission electron microscopy (STEM),  energy dispersive spectroscopy (EDS), high angle annual dark field (STEM-HAADF) spatial mapping. (Appendices A and B) The AgNWs and AgNPLs are transferred by dip-coating onto Silicon substrates with 300 nm oxide, using a Kibron dipper. Subsequently langmuir-schaefer (LS) self\-assembled monolayers are prepared and transferred using a KSV langmuir setup. (Appendix C) The transferred nanostructures are imaged by atomic force microscopy (AFM).

\begin{figure}[htp]
\centering
\includegraphics{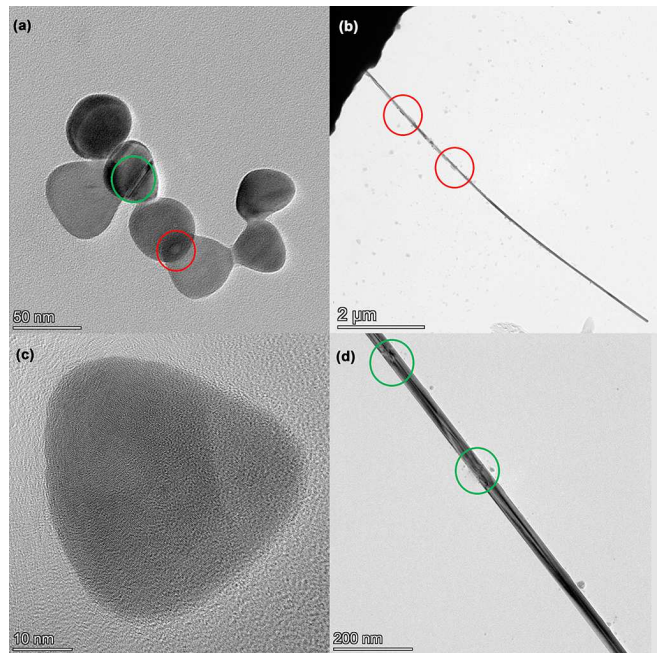}
  \caption{(a) and (b) show the TEM images of Silver nanoplatelets (AgNPLs) and silver nanowires (AgNWs) of aspect ratios of 3 and 241 respectively. (c) and (d) show the high resolution TEM images of Silver nanoplatelets and Silver nanowires. The encircled regions in red and green, show the ridge and groove type defects respectively.}
  \label{fgr:1}
\end{figure}

The absorption spectra of AgNPLs are measured in a Perkin-Elmer lambda-35 solution uv-vis absorption spectrometer. The Photoluminescence (PL) and time resolved photoluminescence (TRPL) are measured with a Witec alpha 300 confocal microscope equipped with a peltier cooled CCD spectrometer and a Picoquant SPAD detector. The PL spectra are excited with a 532 nm diode laser and exciting laser line is cut-off with 532 nm edge long pass filter. The PL integration time is set at 5 s and averaged over such 4 accumulation cycles. The TRPL spectra are excited with a 405 nm pulsed laser and cutoff with a set of 405 nm and 488 nm edge long pass filters. The TRPL integration time is set as 5 s and 10 accumulation cycles.

\section{Results and Discussion}
The TEM images of the chemically synthesized Silver nano-platelets (AgNPLs) and Silver nanowires (AgNWs) are shown in Fig. 1. The chemically synthesized nanostructures generally are defective. The defects are generally grouped into ridge and groove defects.\cite{PhysRevB.83.075433} Such nm sized defects, strongly scatter the surface plasmons.\cite{PhysRevLett.78.4269} The ridge and groove defects scatter the surface plasmons and are  can be used as hot spots\cite{PhysRevB.95.115441} for surface enhanced raman scattering. Such defects can also create shoulders in scattering spectra.

In subsection A, the coupling data of AgNPLs with AQDs is discussed and AgNW-AQD coupling data is discussed in the subsection B.
\subsection{AgNPL-AQD coupling}
The AgNPLs are deposited on Silicon substrates and imaged by AFM. Then 2 monolayers of AQDs are deposited on the AgNPLs directly. The AgNPLs aspect ratio can be controlled by varying the precursor concentration used in synthesis. Three sets of AgNPLs with aspect ratios a\textsubscript{1}=400, a\textsubscript{2}=261 and a\textsubscript{3}=3 are synthesized. Only the AgNPLs with aspect ratios a\textsubscript{1} and a\textsubscript{2} are used for coupling to AQDs. The synthesized triangular AgNPLs edges are blunted. This is known as snipping.\cite{10.1021/jp026731y} With large aspect ratios\cite{10.1021/jp901248e} and snipping of AgNPLs, the in-plane dipolar modes of AgNPLs move to higher wavelengths, typically to infra red region of light.
\begin{figure}[htp]
\centering
\includegraphics{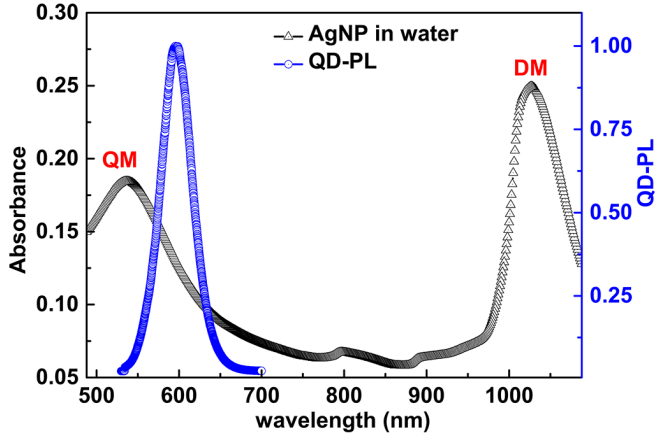}
  \caption{shows the AgNPL (a\textsubscript{3}=3) absorbance and AQD PL superimposed over each other.}
  \label{fgr:5}
\end{figure}
The position of DM modes of AgNPLs of aspect ratio a\textsubscript{3}=3 is at 1027 nm, as shown in Fig. 2. The Absorption spectra of AgNPLs with higher aspect ratios are beyond the detection range of our instrument facilities, so they are not shown.
\begin{figure}[htp]
\centering
\includegraphics{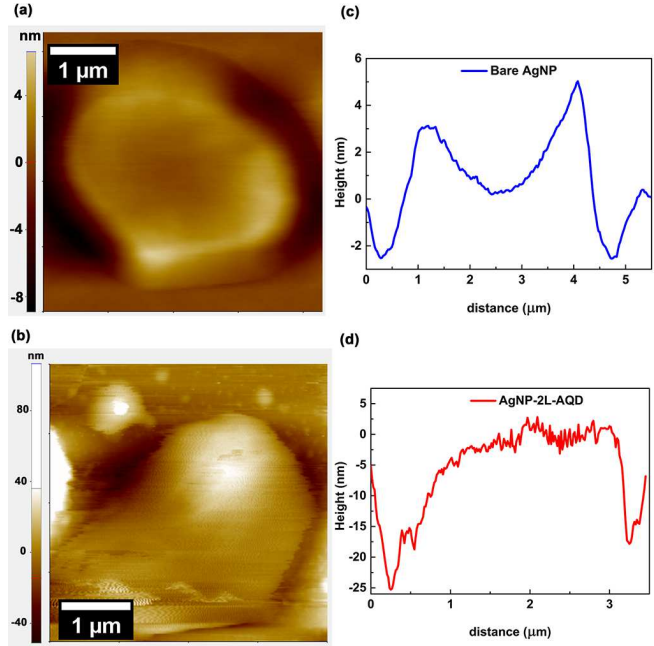}
  \caption{(a) and (b) show the AFM images of a single AgNPL and a single AgNPL coated with 2L-AQDs, respectively. (c) and (d) show the corresponding AFM height profiles of single AgNPL and a single AgNPL coated with 2L-AQDs, respectively}
  \label{fgr:6}
\end{figure}
AgNPLs of two aspect ratios a\textsubscript{1} and a\textsubscript{2} are dip-coated on to Silicon substrates. Then two consecutive AQDs monolayers are deposited on the AgNPLs. The AFM image and topography profile of a single bare AgNPL and 2 AQD monolayers deposited on a single AgNPL are shown in Fig. 3, respectively. The AFM images of the multiple bare AgNPLs and AQD deposited AgNPLs, are shown in Figs. 4(a) and 4(b). The AFM images indicate that the AgNPLs are completely covered with AQDs.
\begin{figure}[htp]
\centering
\includegraphics{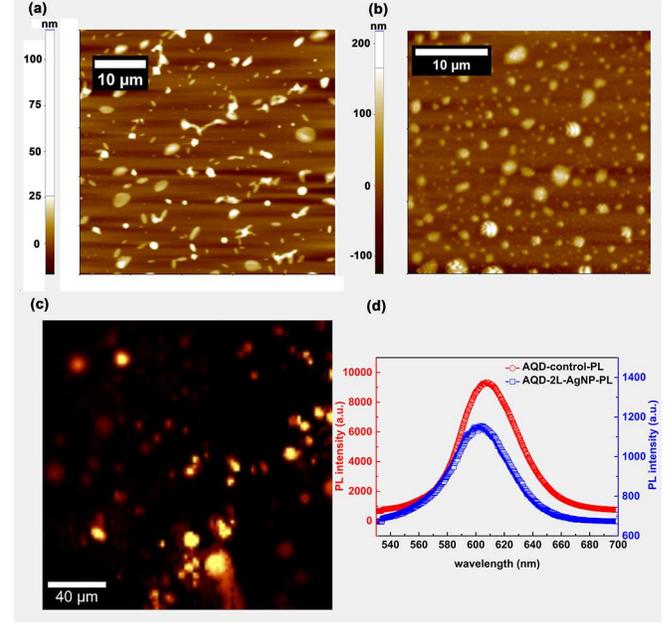}
  \caption{(a) and (b) show the AFM images of bare AgNPLs on Silicon and 2 AQD monolayer coated AgNPLs on Silicon respectively.(c) shows the PL emission spatial map of 2L-AQDs-AgNPLs system. (d) shows the typical PL spectra of of 2L-AQDs and 2L-AQDs-coupled to a single AgNPL.}
  \label{fgr:7}
\end{figure}
The PL spatial map of the coupled AQDs-AgNPLs is shown in Fig. 4 (c). The bright spots in PL spatial map are on AgNPLs. A typical spectra of AQDs coupled to a single AgNPL is shown in Fig. 4(d) along with control PL spectra of 2 monolayer AQDs on Silicon. Similarly the TRPL spectra is collected for control AQDs and AQDs coupled to single AgNPL. The PL spectra is not deformed and shows no extra features as shoulders. The TRPL spectra show that the AQD SE rate is inhibited on the AgNPL and is AQD emission rate faster in case of the control sample.
\begin{figure}[htp]
\centering
\includegraphics{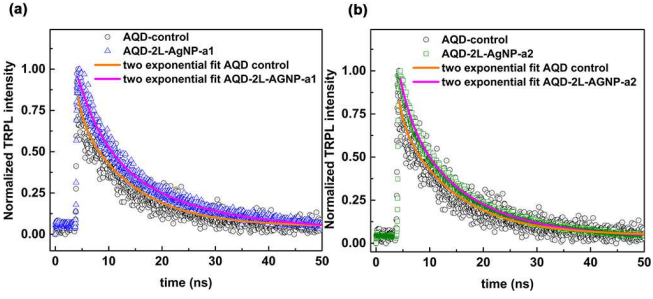}
  \caption{(a) and (b) show the respective TRPL decay profiles of 2L-AQDs on Silicon and 2L-AQDs on single AgNPLs with aspect ratios a\textsubscript{1} and a\textsubscript{2}.}
  \label{fgr:8}
\end{figure}
This confirms that the AQDs are weakly coupled to AgNPL and the coupled system shows suppression of SE.\cite{PhysRevLett.55.67,PhysRevLett.55.2137} To quantify, the SE rate inhibition of AQDs, the TRPL intensity (I) is fitted with the following function, where the A\textsubscript{1} and A\textsubscript{1} are the amplitudes of the decay functions. $\tau_1$ and $\tau_2$ are the respective decay lifetimes. The fitting parameters are shown in Table I.
\[\textup{I}=\textup{A}_1{\cdot}e^{-\frac{\textup{t}}{\tau_1}}+\textup{A}_2{\cdot}e^{-\frac{\textup{t}}{\tau_2}}\tag{5}\]
The weighted average lifetime is given by\[\tau_\textup{avg}=\frac{\textup{A}\textsubscript{1}\cdot\tau_1+\textup{A}\textsubscript{2}\cdot\tau_2}{\textup{A}\textsubscript{1}+\textup{A}\textsubscript{2}}\tag{6}\]
\begin{table}
\caption{The table shows the fitting components of lifetime of AQDs coupled to AgNPL}
\begin{center}
\begin{tabular}{||c c c c c c||} 
 \hline
 Sample & A\textsubscript{1} & $\tau_{1}$ (ns) & A\textsubscript{2} & $\tau_{2}$ (ns) & $\tau_{\textup{avg}}$ (ns)\\ [0.5ex] 
 \hline\hline
 AQD & 0.13 & 1.06 & 0.66 & 10.37 & 8.86 $\pm$ 1.46\\ 
 \hline
 AQD-AgNPL-a1 & 0.37 & 4.86 & 0.54 & 16.74 & 11.89 $\pm$ 1.36\\
 \hline
 AQD-AgNPL-a2 & 0.24 & 2.68 & 0.68 & 12.25 & 9.78 $\pm$ 0.34\\
 \hline
 \end{tabular}
\end{center}
\end{table}
The Purcell factor is calculated from equation (1), using weighted average lifetime from fitting TRPL spectra. The Purcell factors (F\textsubscript{P}) for single AgNPL of aspect ratio a\textsubscript{1} and a\textsubscript{2} are 0.75 $\pm$ 0.21 and 0.96 $\pm$ 0.18, by fitting the TRPL decay prfiles in Fig. 5. As a\textsubscript{1} $>$ a\textsubscript{2}, the DM wavelength of AgNPL a\textsubscript{1} is larger than that of AgNPL a\textsubscript{2}. So the Purcell factor of AQDs coupled to AgNPL a\textsubscript{1} is smaller, which is indicator of increased inhibition of AQD SE. So with increasing aspect ratio of AgNPLs, Purcell inhibition of SE also increases.

\subsection{AgNW-AQD coupling}
The control sample is bare Silver nanowires (AgNWs) deposited on the Silicon. The control sample is imaged by TEM and AFM. The AFM imaging indicates that the AgNW thickness is approximately 36.1 nm, as shown in Fig. 6(c). The typical aspect ratio of AgNWs is about 240. On a separate control sample, 10 nm of spacer layer is deposited onto the nanowires by spin-coating 5 mg/ml polymethyl methacrylate (PMMA) solution in toluene. (Appendix C) The one monolayer (1L) of AQDs is transferred by the LS method on top of the spacer and the AQD monolayer is about 5.6 nm thick.\cite{10.48550/arxiv.2212.13510} The AFM height profile of the AQD deposited sample with spacer, is 55.8 nm as shown in Fig. 6(d).

\begin{figure}[htp]
\centering
\includegraphics{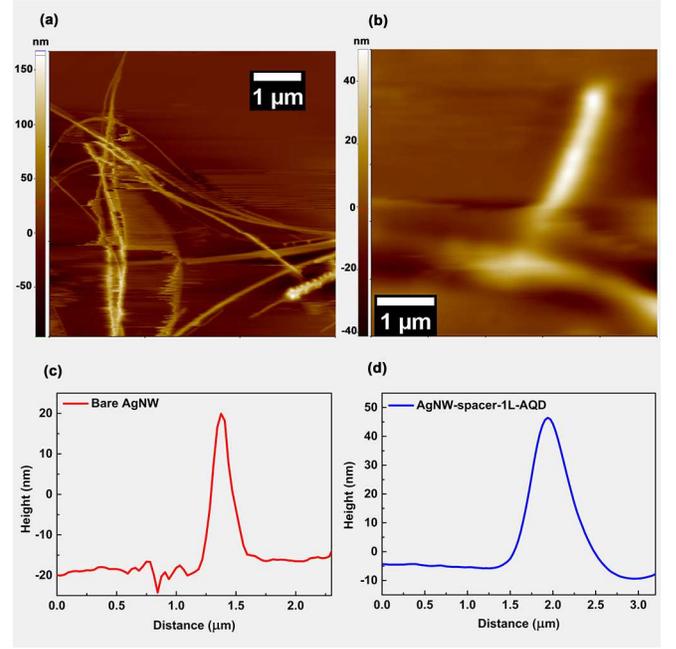}
  \caption{(a) and (b) show the AFM images of bare AgNWs and 1L AQDs coated AgNW with 10 nm polymer spacer, respectively.(c) and (d) show the AFM profiles of a bare AgNW and 1L AQDs coated AgNW, with 10 nm polymer spacer respectively.}
  \label{fgr:2}
\end{figure}

The PL spectra of the coupled AQD-AgNW system are measured. The optical image of the system is shown in Fig. 7(a) and the spatial PL emission map is shown in Fig. 7(b). The PL spectra of control sample: monolayer AQD on spacer and single AgNW-spacer-AQD monolayer are shown in Fig. 7(c). The PL spectra of the coupled system shows a shoulder feature. The PL spectra is fitted with two Gaussian functions and the peak to peak separation is 160 meV. The amplitude and full width half maximum (FWHM) of Gaussian functions 1 and 2 are represented by A\textsubscript{1}, A\textsubscript{2}, w\textsubscript{1}, w\textsubscript{2} respectively.
\[\textup{I}(\lambda)= \textup{I}_\textup{o} + \textup{A}_{1}{\cdot}e^{-\big(\frac{\lambda-\lambda_{1}}{2\textup{w}_1^2}\big)} + \textup{A}_{2}{\cdot}e^{-\big(\frac{\lambda-\lambda_{2}}{2\textup{w}_2^2}\big)} \tag{7}\]
\[\textup{w}_\textup{avg}=\frac{\textup{A}_{1}\cdot\textup{w}_1+\textup{A}_2\cdot\textup{w}_2}{\textup{A}_1+\textup{A}_{2}}\tag{8}\]
The PL fitting parameters are indicated in Table II. Here $\lambda_1$ and $\lambda_2$ are the central positions of the constituent gaussian functions 1 and 2.
\begin{table}
\caption{The table shows the fitting components of lifetime of AQDs on spacer and AQDs coupled to spacer coated AgNWs}
\begin{center}
\begin{tabular}{||c c c c c c||} 
 \hline
 Sample & A\textsubscript{1} & $\textup{w}_{1}$ & A\textsubscript{2} & $\textup{w}_{2}$ & w\textsubscript{avg}\\ [0.5ex] 
 \hline\hline
 1L-AQD-AgNW & 0.21 & 22.19 & 0.95 & 47.22 & 42.67 $\pm$ 0.24\\
 \hline
 5L-AQD-AgNW & 0.18 & 23.91 & 0.94 & 56.40 & 51.27 $\pm$ 0.38\\
 \hline
\end{tabular}
\end{center}
\end{table}
The $\lambda_1$ and $\lambda_2 $  positions for multiple AgNWs are 554.55 nm $\pm$ 0.11 nm and  597.97 nm $\pm$ 0.04 nm. For single AgNW, they are 553.60 nm $\pm$ 0.08 nm and 596.45 nm $\pm$ 0.03 nm.
Such shoulders in quantum dot PL spectra might be observed due to Rabi splitting of strongly coupled systems.\cite{10.1038/s41467-018-06450-4} For N emitter-strong coupling, the Rabi splitting scales with number of emitters.\cite{10.1021/nn100585h} The coupling coefficient (g) of N dipolar emitters with a cavity mode field (E) is given by\cite{PhysRevA.23.3107,10.1098/rsta.2010.0333}
\[\hbar\textup{g}=\sqrt{\textup{N}}\big(\vec{\mu}\cdot\vec{\textup{E}}\big)\tag{9}\]
Here $\mu$ is transition dipole moment of the emitter.\cite{10.1119/1.12937} The Rabi splitting magnitude is 2g. So the Rabi splitting, i.e., the peak to peak separation should scale with number of emitters. So if the number of emitters is increased by 5 times, then Rabi splitting magnitude should increase 2.2 fold.
\begin{figure}[htp]
\centering
\includegraphics{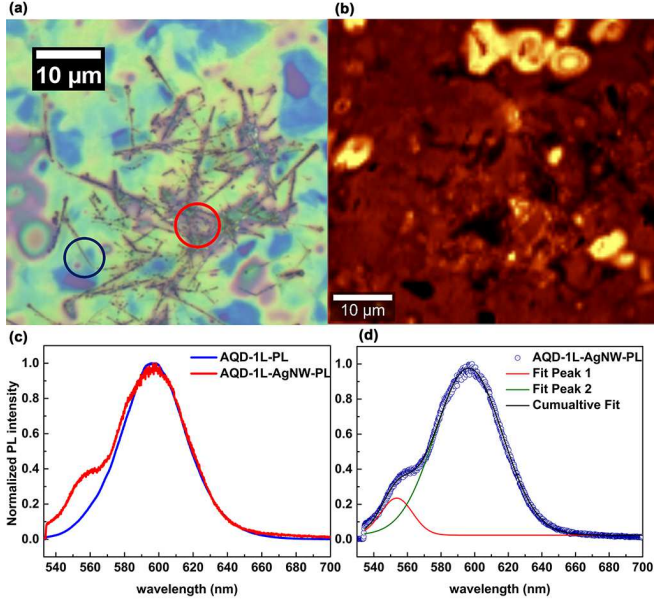}
  \caption{(a) shows the optical image of the 1L AQDs deposited on 10 nm spacer coated AgNWs on Silicon. A typical single nanowire is shown by the blue circle. Multiple nanowire clusters are shown by the red circle. (b) shows the corresponding PL spatial map of the 1L AQDs deposited on 10 nm spacer coated AgNWs on Silicon. (c) shows the PL spectral deformation of 1L-AQDs-coupled to AgNWs relative to 1L-AQDs.(d) shows the fitting of PL spectra of 1L-AQDs-coupled to AgNWs.}
  \label{fgr:3}
\end{figure}
5 monolayers of AQDs are transferred on an AgNW with spacer and compared with 1 monolayer AQDs on a AgNW. The PL spectra of both the samples are measured. The shoulder separation ($\lambda_2-\lambda_1$) for 1L AQDs coupled to AgNW is 42.85 nm. The separation should scale to 95.82 nm for 5L AQDs coupled to AgNW, in case of true strong coupling. Instead the shoulder separation of the 5L AQDs coupled to AgNW is measured as 43.42 nm, which is with in 1\% change from the initial value for 1L AQDs coupled to AgNW. This null result confirms that there is no emitter number dependence. So, strong coupling is conclusively ruled out. The shoulder is attributed to scattering by AgNW.

Also, as the AQD concentration changes from 1L to 5L, the weighted average spectral line width (w\textsubscript{avg}) increases from 42.67 nm to 51.27 nm, i.e. by 8.60 nm $\pm$ 0.62 nm. The PL spectral broadening with increased emitter concentration is indicative of weak coupling. To find out more about the nature of coupling of AQDs to AgNWs radiative lifetime of AQDs is measured.
\begin{figure}[htp]
\centering
\includegraphics{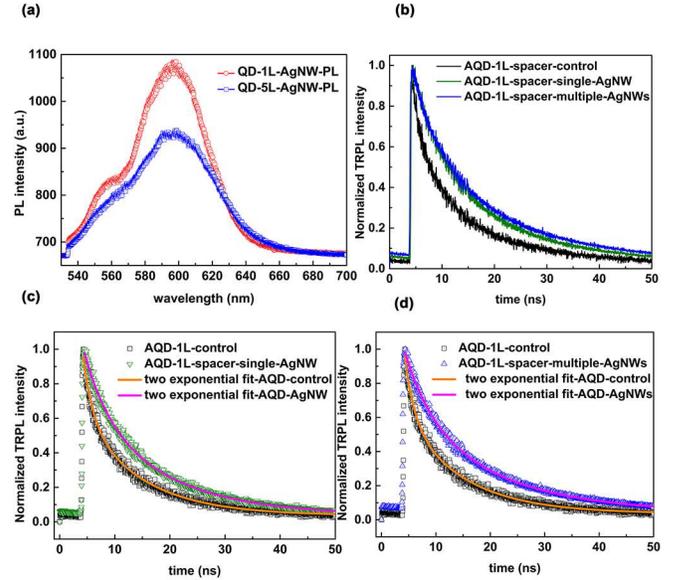}
  \caption{(a) shows the PL spectra of 1L-AQDs-coupled to AgNWs and 5L-AQDs-coupled to AgNWs. (b) shows the normalized raw SE rate profiles of 1L-AQDs on spacer coated Silicon, 1L-AQDs-spacer-single AgNW and 1L-AQDs-spacer-multiple AgNWs respectively.(c) and (d) show the inhibition of SE rate of 1L-AQDs for single and multiple AgNWs respectively.}
  \label{fgr:4}
\end{figure}
It is evident from the raw data in Fig. 8(b), that the SE rate of 1L-AQDs on AgNWs is slower compared to free space decay rate of 1L-AQDs, which confirms the coupled system is in weak coupling. Similar fitting approach of AgNPLs is adapted for AgNWs. The fitting parameters are shown in Table III.

\begin{table}
\caption{The table shows the fitting components of lifetime of AQDs on spacer and AQDs coupled to spacer coated single and multiple AgNWs}
\begin{center}
\begin{tabular}{||c c c c c c ||} 
 \hline
 Sample & A\textsubscript{1} & $\tau_{1}$ (ns) & A\textsubscript{2} & $\tau_{2}$ (ns) & $\tau_{\textup{avg}}$ (ns)\\ [0.5ex] 
 \hline\hline
AQD-1L & 0.35 & 1.77 & 0.57 & 10.78 & 7.35 $\pm$ 0.07 \\
\hline
AQD-1L-AgNW & 0.29 & 4.05 & 0.65 & 14.12 & 10.99 $\pm$ 0.35\\
\hline
AQD-1L-AgNWs & 0.31 & 4.57 & 0.63 & 16.02 & 12.24 $\pm$ 0.84\\
\hline
\end{tabular}
\end{center}
\end{table}
The Purcell factor (F) is calculated from the ratio of of AQD lifetime to the weighted average life time ($\tau_{\textup{avg}}$). For Single AgNWs, the Purcell factor is 0.67 $\pm$ 0.06 and for multiple AgNWs it is 0.60 $\pm$ 0.12.

The SE inhibition of AQDs emitting in the visible region of light is expected, as the plasmonic resonances of Silver nanowires are either in ultraviolet or infrared regions.\cite{10.1021/acs.jpcc.2c01251,10.1021/nl200634w} The lifetime decay profiles reaffirm that the absence of resonant cavity modes inhibits the SE of AQDs.\cite{10.1063/1.4818131}
\section{Conclusion}
The Silver nanowire and Silver nanoplatelet plasmonic resonances are off resonant to alloyed quantum dot spontaneous emission frequency. This results in inhibition of spontaneous emission of quantum dots coupled to Silver nanowires and Silver nanoplatelets. The spontaneous emission inhibition is quantified in terms of Purcell factor. Purcell factors of 0.64 and 0.79 are observed for quantum dots coupled to silver nanowire and silver nanoplatelet respectively. With increasing aspect ratio of AgNPL and with increasing number of silver nanowires, the spontaneous emission rate is increasingly inhibited.

\begin{acknowledgments}
 B. Tongbram thanks the Department of science and technology (DST), Inspire faculty programme for fellowship. H.R. Kalluru thanks the Micro and nano characterization facility (MNCF-CeNSE), IISc for access to titan themis 300 kV TEM facility.
\end{acknowledgments}
\appendix
\section{AgNPL synthesis method and Characterization}
The Silver Nanoplates (AgNPLs) are synthesized in hydrophilic phase, with capping agent Polyvinylpyrrolidone (PVP), as per the reported protocol.\cite{10.1021/ja2080345} The glassware used for the synthesis is cleaned following the RCA SC-I protocol, subsequently rinsed thrice in DI water and dried. Silver nitrate (AgNO\textsubscript{3}-99\%), sodium borohydride (NaBH\textsubscript{4}-99.99\%), sodium tri{-}citrate dihydrate (TSCDH-99\%) and polyvinylpyrrolidone-40K are procured from Merck. 30\% w/W hydrogen peroxide solution is purchased from SDFCL.

A 100 ml borosilicate conical glass flask and 46.68 ml DI water is added to the flask. Then 120 $\mu$L of hydrogen peroxide solution is added to the conical flask. 140 mg of PVP is dissolved in 1 ml DI water and the whole PVP solution is added to the conical flask. 22.3 mg of TSCDH is dissolved in 1 ml DI water and the solution is added to the flask. Now the flask is placed on magnetic stirrer setup. The stirrer was turned on and set at 800 rpm for rigorous mixing of precursors, at room temperature (300 K).

A glass vial is placed in an ice bath and 4 ml DI water is added to it. The ice cold water is kept ready for dissolving NaBH\textsubscript{4}. 8.6 mg silver nitrate is dissolved in 1 ml DI water and 0.2 ml of the solution is added to the flask. Immediately the solution colour turned to pale yellow. 15.1 mg of NaBH\textsubscript{4} is added to ice cold water and mixed thoroughly using a suction pipette. NaBH\textsubscript{4} solution needs to be immediately used after preparation, as it degrades with time. 1 ml of NaBH\textsubscript{4} solution is added to the mixture of precursors. The reaction mixture immediately turned deep brownish yellow.

After approximately 30 minutes of continuous stirring, the solution turns dark red and finally to deep brown. The whole colour change process happens with in 1-2 minutes. The colour change of solution is due to shifting of localised surface plasmon resonance peak shift due to lateral growth of AgNPLs in the reaction mixture. The reaction mixture is cleaned by centrifuging at 10000 rpm for 10 minutes. The precipitate is dark in colour and the supernatant is brownish. The supernatant is discarded and precipitate is again dispersed in DI water. Then the centrifuging process is repeated further twice, by selecting precipitate and discarding supernatant. After third centrifuging, the precipitate is dispersed in ethanol or DI water. The resultant solution is purple and is used for further characterization and measurements.

For TEM measurements, the AgNPL solution (20 $\mu$g/ml) in ethanol is drop casted on to a copper transmission electron microscopy (TEM) grid and dried in a dessicator under vacuum for 12 hours. The dried TEM grid is cleaned with argon plasma (chamber vacuum 5x10\textsuperscript{-4} Torr and incident power 22 W) for 40 seconds. STEM-HAADF mapping of the AgNPLs is shown in Fig. 9. The Silver characterstic X-ray intensity follows the contours of the AgNPL volume. This indicates that the AgNPLs are made of Silver. The Carbon X-ray intensity map indicates the distribution of PVP ligand over the AgNPL. The AgNPL edge length and thickness histograms are shown in Fig. 10(c) and 10(d) respectively.

\begin{figure}[htp]
\centering
\includegraphics{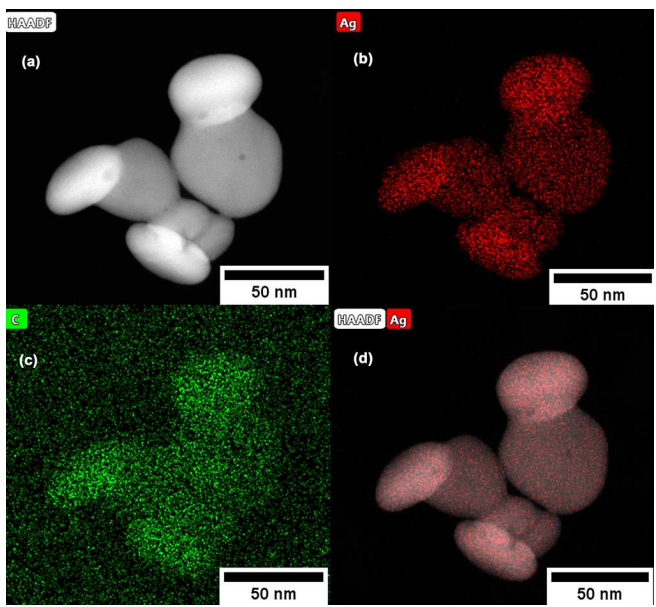}
  \caption{(a) shows the typical dark field STEM-HAADF images of AgNPLs on a TEM grid. (b) and (c) show the silver and carbon atomic distribution on AgNPLs of aspect ratio a\textsubscript{3}. (d) shows the superimposed silver atomic distribution and darkfield STEM-HAADF image.}
  \label{fgr:9}
\end{figure}

\begin{figure}[htp]
\centering
\includegraphics{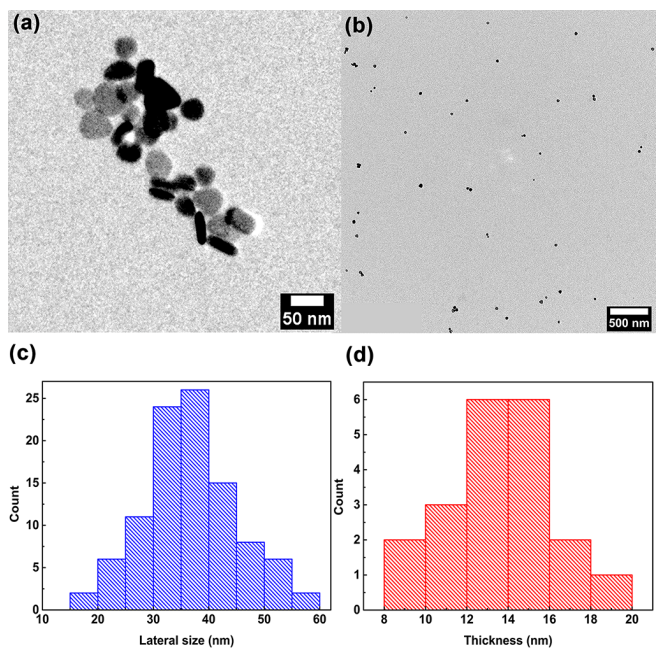}
  \caption{(a) and (b) show the typical magnified and large area view of the TEM images of drop casted AgNPLs of aspect ratio a\textsubscript{3} on TEM grid.(c) and (d) show the distribution in lateral size and thickness of AgNPLs of aspect ratio a\textsubscript{3}}
  \label{fgr:10}
\end{figure}

The AgNPLs of three aspect ratios are synthesized by making the following changes. The AgNPLs of aspect ratio a\textsubscript{1} are synthesized by following above mentioned procedure. For synthesizing the AgNPLs with aspect ratio a\textsubscript{2}, the volumes of the precursors (AgNO\textsubscript{3} ; NaBH\textsubscript{4}) dropped in the conical flask are changed from (0.2 ml ; 1 ml) to (0.1 ml; 0.5 ml) respectively. To synthesize the AgNPLs of aspect ratio of a\textsubscript{3}, the reaction volume of precursors required for AgNPLs of aspect ratio of a\textsubscript{2} is reduced by half, such that the reaction mixture volume is 25 ml.

\begin{table}
\caption{The table shows the synthesized nanostructure specifications.}
\begin{center}
\begin{tabular}{||c c c c||} 
 \hline
 Sample & Lateral size ($\mu$m) & thickness (nm) & Aspect ratio\\ [0.5ex] 
 \hline\hline
 AgNW & 8.435 & 35 & 241\\ 
 \hline
 AgNP-a1 & 2.243 & 5.6 & 400\\
 \hline
 AgNP-a2 & 1.176 & 4.5 & 261\\
 \hline
AgNP-a3 & 0.043 & 14 & 3\\
 \hline
\end{tabular}
\end{center}
\end{table}

\section{AgNW synthesis method and Characterization}
The Silver nanowires (AgNWs) are synthesized in hydrophobic phase with capping agent Oleyl amine (OAm), as per the reported protocol.\cite{10.1039/C5RA13884A} The glassware used for the synthesis is cleaned by the RCA SC-I protocol and rinsed thrice in DI water and dried. Silver bromide (AgBr-99\%), copper chloride (CuCl\textsubscript{2}-99.5\%), n-Hexane (99\%) and oleyl amine (70\%) are procured from Merck. 0.3 gL\textsuperscript{-1} CuCl\textsubscript{2}{-}OAm solution is prepared by dissolving 3 mg of CuCl\textsubscript{2} in 10 mL of OAm at $\textup{60}^{\textup{o}}$ C. The temperature is maintained for 10 min and then the solution is cooled to ambient temperature.

In a round bottom borosilicate glass flask 0.1g AgBr, 34 $\mu$L of CuCl\textsubscript{2}-OAm solution and 5 ml OAm is added. The flask is heated to $\textup{160}^{\textup{o}}$ C and maintained at the same temperature for 6 hours. Then the heating element is turned off and reaction mixture is allowed to reach ambient temperature naturally. The final reaction mixture colour is dark gray, indicating formation of AgNWs.

The reaction mixture is dispersed in 15 ml n-hexane and centrifuged at 6000 rpm for 10 minutes. The precipitate is then dispersed in hexane and centrifuged for two more cycles. The precipitate after 3 cycles is dispersed in n-hexane and stored in dark for characterization and measurements.

\begin{figure}[htp]
\centering
\includegraphics{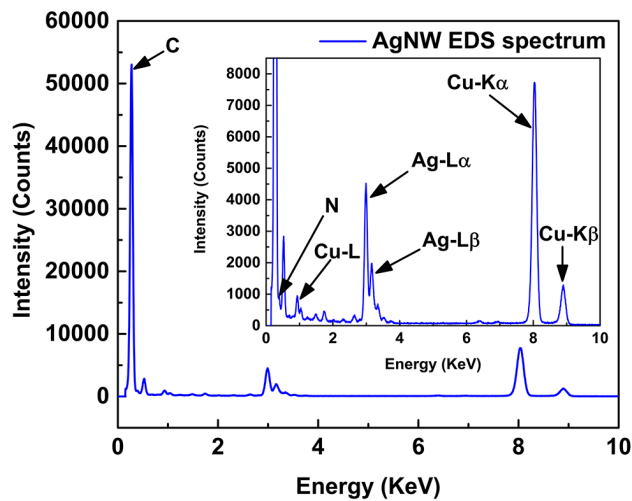}
  \caption{shows the AgNW EDS spectra with characteristic X-ray peaks of constituent atoms. The inlay shows the zoomed peaks of the EDS spectra of AgNWs.}
  \label{fgr:12}
\end{figure}

For TEM measurements, the AgNW solution (10 $\mu$g/ml) in hexane is drop casted on to a copper transmission electron microscopy (TEM) grid and processed identical manner of AgNPLs on TEM grid. The EDS spectra of the AgNWs is measured and shown in Fig. 11. The elemental distribution is shown in table IV. The carbon and nitrogen content is attributed to oleyl amine ligands. The Copper content is attributed to both Copper TEM grid and CuCl\textsubscript{2} seeding process of AgNWs. The Silver content is attributed to AgNWs.

\begin{table}
\caption{The table shows the elemental analysis obtained from the EDS spectrum of AgNWs.}
\begin{center}
\begin{tabular}{||c c c||} 
 \hline
 Element & Atomic fraction (\%) & Error (\%)\\ [0.5ex] 
 \hline\hline
 Carbon & 95.79 & 4.05\\ 
 \hline
 Nitrogen & 1.39 & 0.28\\
 \hline
 Copper & 0.83 & 0.11\\
 \hline
Silver & 1.98 & 0.24\\
 \hline
\end{tabular}
\end{center}
\end{table}

\section{Sample preparation procedure}
The synthesized AgNPLs are hydrophilic and AgNWs are hydrophobic in nature. AgNPLs are transferred onto Silicon substrate by dip-coating at water and hexane interface, by following the procedure mentioned in report.\cite{10.1038/ncomms7990}  The self assembly of hydrophilic particles at water-hexane interface reduces clustering of AgNPLs and is proffered for studying properties of single AgNPLs. Typically in a large glass petridish, the substrate is placed and is attached to a motorized dipper (KSV make) and 200 ml deionized (DI) water is poured over the substrate, till the substrate gets completely immersed. 
\begin{figure}[htp]
\centering
\includegraphics{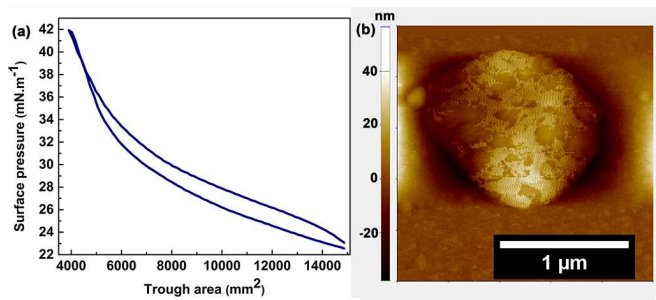}
  \caption{shows the typical isothermal compression and compression cycle for transfer of AQDs with a KSV-Nima LS setup.(b) shows the AFM image of an AgNPL with aspect ratio a\textsubscript{2}=261.}
  \label{fgr:13}
\end{figure}
Then 200 ml hexane is poured over DI water, which floats over water. Then 3 ml of 2 mg/ml solution of AgNPLs in DI water is dispersed in the DI water phase gently below Hexane. The system is allowed to stabilize for 15 minutes. There are two interfaces in this system. First interface is at the hexane-air boundary. The other interface is the hexane-water boundary.

Then the substrate is brought up from water phase to the hexane-water interface, at a rate of 10 mm/min. Once the substrate touches water-hexane interface, the AgNPLs at the hexane-water interface gets transferred onto the substrate. The substrate is then moved up to hexane-air interface and the hexane is allowed to dry at ambient conditions.

AgNWs are transferred onto Silicon substrate by dip-coating at hexane air interface in a similar procedure followed for AgNPLs, with a major difference. The amount of hexane poured over DI water is limited to 10 ml, so that a thin Hexane phase floats over DI water. Then 5 ml of 2.2 mg/ml AgNW solution in hexane is dispersed in DI water phase. AgNWs are hydrophobic and move immediately to hexane phase. The system is allowed to stabilize for 15 minutes and then the substrate is brought up towards the hexane-water interface at a rate of 10 mm/min. As soon as the substrate crosses hexane-air interface, the AgNWs get transferred on the substrate. The substrate is then dried at ambient conditions.

Poly-methyl-methacryalate (PMMA), 350K molecular weight is procured from Merck and 5 mg/ml solution is prepared in Toluene. The solution is spin-coated on a clean Silicon substrate at 3000 rpm and for 60 s. The resultant film thickness is characterized by X-ray reflectivity (XRR) measurement. The XRR fringe separation is a measure of  film thickness (h). The film thickness is calculated as per procedure mentioned in\cite{10.1016/0040-6090(93)90499-F}

The film thickness turns out as 9.45 nm. Monolayers of AQDs are synthesized as per the reported procedure.\cite{10.1007/s11144-014-0813-0,10.1021/jp800063f} The AQDs are transferred by self-assembly via Langmuir-Schaefer (LS) method,\cite{10.1016/j.ccr.2013.07.023} as per the procedure mentioned in the report.\cite{PhysRevB.100.155413,10.48550/arxiv.2212.13510} The monolayer formation in a typical LS method is indicated by saturation of surface pressure in isothermal compression cycle of the AQDs\cite{10.1021/la904474h}. The typical compression and expansion cycle of LS isotherm is shown in Fig. 12(a). The formed monolayer is transferred on to AgNWs with spacer, by lowering and stamping the substrate onto the self assembled monolayer. Once the stamping is done, the substrate is retracted and dried in ambient conditions.

\bibliography{apssamp}
\bibliographystyle{apsrev4-2}
\end{document}